\documentclass[aps,prl,reprint,showpacs]{revtex4-1}

\usepackage{graphicx}
\usepackage{dcolumn}
\usepackage{bm}
\usepackage{epsfig}
\usepackage{amsmath}
\usepackage[usenames,dvipsnames]{color}
\usepackage{booktabs}
\begin{document}

\author{Mahmoud M. Asmar}
\email{asmar@phy.ohiou.edu} \affiliation{Department of Physics and
Astronomy and Nanoscale and Quantum Phenomena Institute, Ohio
University, Athens, Ohio 45701-2979, USA} \affiliation{Dahlem Center
for Complex Quantum Systems and Fachbereich Physik, Freie
Universit\"at Berlin, 14195 Berlin, Germany}
\author{Sergio E. Ulloa}
\email{ulloa@ohio.edu} \affiliation{Department of Physics and
Astronomy and Nanoscale and Quantum Phenomena Institute, Ohio
University, Athens, Ohio 45701-2979, USA} \affiliation{Dahlem Center
for Complex Quantum Systems and Fachbereich Physik, Freie
Universit\"at Berlin, 14195 Berlin, Germany}

\title{Spin Orbit Interaction and Isotropic Electronic Transport in Graphene}

\begin{abstract}

Broken symmetries in graphene affect the massless nature of its
charge carriers. We present an analysis of scattering by defects in
graphene in the presence of spin-orbit interactions (SOIs). A
characteristic constant ratio ($\simeq 2$) of the transport to
elastic times for massless electrons signals the {\em anisotropy} of
the scattering. We show that SOIs lead to a drastic decrease of this
ratio, especially at low carrier concentrations, while the
scattering becomes increasingly {\em isotropic}. As the strength of
the SOI determines the energy (carrier concentration) where this
drop is more evident, this effect could help evaluate these
interactions through transport measurements.

\end{abstract}

\pacs{72.10.Fk, 75.76.+j, 72.80.Vp, 03.65.Pm}

\maketitle

The discovery of graphene has stimulated numerous theoretical and
experimental works \cite{novo}, opening  new doors for promising new
technology due to its low dimensionality and high carrier mobility.
The low energy electron dynamics is described by two inequivalent
points at the Brillouin zone ($K$ and $K^{'}$) known as Dirac
points, since the linear dispersion is equivalent to two-dimensional
massless Dirac fermions \cite{elctronincproperties,seminov}.

The importance of graphene on transport devices also motivates the
identification and understanding of spin dynamics \cite{coherence},
as an important element in the development of spintronics. In
graphene, interface or bulk broken symmetries allow for the
existence of two kinds of spin orbit interaction (SOIs) that affect
spin dynamics in different ways \cite{quntums}. The hexagonal
arrangement of carbon atoms allows an \emph{intrinsic} SOI that
respects lattice symmetries and can be seen to arise from the atomic
SO coupling. This generates a gap in the spectrum, a mass term in
the Dirac equation with sign depending on the spin, pseudospin and
Dirac valley \cite{intrinsic,intrinsic1}. An inversion asymmetry in
graphene could also generate an \emph{extrinsic} Rashba SOI,
resulting from the effect of substrates, impurities generating
$sp^{3}$ distortions--such as hydrogen, fluorine or
gold--perpendicular electric fields, or lattice corrugations
\cite{experiment1,impurityso,golddep,curvature2,curvature3,Nii}.
Intercalation of gold under graphene deposited on nickel substrates
results in very large Rashba interactions \cite{Nii}, while a large
enhancement was observed in weakly hydrogenated samples
\cite{colossal}. In addition, recent theoretical studies have shown
that decoration of graphene with heavy atoms such as indium and
thallium will result in the enhancement of an intrinsic-like SOI in
graphene and the associated quantum spin Hall state \cite{indium}.

Adsorbed impurities \cite{adsorbh,hydrogen}, as well as lattice
vacancies and other local defects in the lattice \cite{vacancies}
provide natural short-range scattering centers known as {\em
resonant scatterers}. Sources of resonant scatterers are also
organic groups~\cite{realistic}, clusters of impurities
\cite{cluster}, or even artificially controlled metallic islands
deposited on the surface of graphene \cite{islands}. Extensive work
has identified the existence of resonant scatterers as the main
mechanism limiting carrier mobility in graphene samples
\cite{resonant,resonant1,realistic}. These conclusions are supported
by the insensitivity to screening effects provided by the different
substrates used \cite{permitiv,permitiv1}, by the independence of
the ratio of the transport to elastic times to the carrier
concentration \cite{helen}, and by the universal presence of the
Raman $D$ peak in graphene devices and its stability after
high-temperature annealing of samples \cite{raman,raman1}.
Experiments performed by Monteverde {\em et al}.\ \cite{helen} used
the transport ($\tau_{tr}$) and elastic ($\tau_{e}$) scattering
times extracted from magnetotransport measurements to probe the
nature of the impurities in single and bilayer graphene. The ratio
of these two characteristic times, $\xi={\tau_{tr}}/{\tau_{e}}$,
describes at low Fermi energies (low carrier concentration) the
degree of angular anisotropy of the scattering process, offering an
interesting insight on the type of impurities present in samples.
One should comment that other work argues that carrier mobility in
graphene is mainly limited by long range scattering from charged
impurities \cite{chrgeds,chrgeds2,chrgeds3,chrgeds4}, also related
to the formation of electron-hole puddles
\cite{elctronincproperties,chrgeds2,pudle,permitiv}.

Short range scatterers are categorized according to the total cross
section, $\sigma_{t}$, they produce \cite{helen}:  ``Small cross
section scatterers'' have $\sigma_{t}\propto k$, where $k$ is the
carrier Fermi wave number ($k \propto E_F \propto \sqrt{n_c}$, with
$n_c$ the carrier density). ``Medium cross section scatterers'' are
referred in the literature also as resonant scatterers, and display
a different dependence, $\sigma_{t}\propto 1/(k\ln^{2}k)$. Finally,
the ``large total cross section scatterers'' or ``unitary'' are
associated with the presence of a long-lived quasibound
state \cite{quasi,Matulis}, and exhibit $\sigma_{t}\propto 1/k$. An
important common property shared {\em by all these regimes} is that
the ratio of the transport to elastic times is determined fully by
the conservation of pseudo-helicity, leading to a value of $2$ at
low energies, as we will discuss below.

We will show that the presence of SOIs leads to an important
transformation of scattering processes in graphene, from highly
anisotropic (zero backscattering) to more or fully isotropic at low
energies, depending on the strength of these interactions. We show
that the Rashba SOI results in the appearance of new unitary
resonances for short-range scatterers, whenever Rashba coupling is
comparable to the Fermi energy. Moreover, we show that the three
different types of short range scatterers (off resonant, resonant,
and unitary), lead to processes with different levels of angular
isotropy, unlike the case with no Rashba SOI when {\em all} short
range scatterers display similar anisotropy. These findings suggest
that transport experiments performed at low carrier concentration
could unveil the local enhancement of the Rashba interaction
produced by impurities, lattice corrugations, or substrate effects,
and provide a direct measurement of its strength.

We consider the presence of intrinsic SOI, affecting the carriers
throughout the graphene system, while an extrinsic scatterer
generates a local potential obstacle and corresponding Rashba SOI;
the Hamiltonian for this system close to the Dirac points is then
given by
\begin{equation}\label{eq1}
H=H_{o}+H_{V}+H_{SO}+H_{R}\; ,
\end{equation}
where $H_{o}=\hbar v_{F}(\tau_{z}\sigma_{x}p_{x}+\sigma_{y}p_{y})$
describes Dirac fermions in graphene,
$H_{SO}=\Delta_{SO}\sigma_{z}\tau_{z}s_{z}$ is the intrinsic SOI,
$H_{V}=V\Theta(R-r)$ is the scattering potential characterized by
strength $V$ over a region $r<R$, and
$H_{R}=\lambda_{R}(\tau_{z}\sigma_{x}s_{y}-s_{x}\sigma_{y})\Theta(R-r)$
is the Rashba SOI \cite{quntums} over the same region; here $\hbar
v_{F}\simeq~6.6$eV$\cdot$\AA, while $\sigma_{\mu}$ and $s_{\mu}$ are
Pauli matrices representing the electron pseudospin $(A,B)$ and spin
$(\uparrow,\downarrow)$, respectively, and $\tau_{z}=\pm1$
identifies the $K$ or $K'$ valleys. $\Delta_{SO}$ and $\lambda_{R}$
are the strengths of intrinsic and Rashba interactions, and $\Theta$
is the Heaviside function. The characteristic size of the scatterers
is assumed to be much larger than the lattice spacing in graphene
for the continuum Dirac description of graphene to be appropriate,
and to neglect intervalley scattering
\cite{elctronincproperties,oldref}.

The analytical form of the spinors \cite{SuppInfo} allows one to use
a partial wave decomposition to study the scattering of an incoming
flux of electrons along the $x$-direction \cite{elastic}, which
takes the asymptotic form away from the scattering center
\begin{equation}\label{plane}
\psi\approx e^{ikr\cos\theta} \,
\chi_{in}+\hat{f}(\theta)\frac{e^{ikr}}{\sqrt {r}} \, \chi_{in} \, ,
\end{equation}
where
$\chi_{in}=(c_{1}\mid\uparrow\rangle,c_{2}\mid\downarrow\rangle)^T$
is a spinor describing the spin weights of the incoming flux with
$|\chi_{in}|^{2}=1$, $k=\sqrt{E^{2}-\Delta_{SO}^{2}}/\hbar v_{F}$,
and $\hat{f}(\theta)$ is a matrix containing the different
scattering amplitudes. The conservation of total angular momentum
$J_{z} = L_z + \hbar \tau_z \sigma_z/2 + \hbar s_z/2$, where $J_z
\psi_n = \hbar n \psi_n$ \cite{SuppInfo}, allows consideration of separate
partial wave components of the incoming wave with a given spin $s$,
$\psi_{n}^{(-)}|s\rangle$. Hence, the full wave function away from
the scattering area is given by
\begin{equation}\label{out}
\psi^{out}_{n}(r,\theta)=\psi_{n}^{(-)}|s\rangle+ \sum_{s'}
S_{n,ss'} \, \psi_{n}^{(+)}|s'\rangle \, ,
\end{equation}
where $s,s'=\uparrow, \downarrow$  and $\psi_n^{(+)}$ is an outgoing
wave. The asymptotic form of the Henkel functions and the
Jacoby-Anger expansion \cite{SuppInfo}, allows one to relate the
wave functions in (\ref{plane}) and (\ref{out}), and characterize
the scattered part of the wave function as $\bar{s}=-s$
\begin{equation}
\psi^{sct}_{n}=\frac{e^{-i\pi/4}}{\sqrt{2\pi k}}
\left(\left(S_{n,ss}-1\right)\psi_{n}^{(+)}|s\rangle + i \bar{s}
S_{n,s\bar{s}} \, \psi_{n}^{(+)}|\bar{s}\rangle\right) \, ,
\end{equation}
leading to the scattering amplitude matrix
\begin{equation}\label{fmatrix}
\hat{f}(\theta)=\frac{e^{-i\pi/4}}{i\sqrt{2\pi k}}\sum_n
{\left[\begin{array}{cc}
                  f_{n,\uparrow\uparrow} & f_{n,\downarrow\uparrow} \\
                  f_{n,\uparrow\downarrow}&f_{n,\downarrow\downarrow}
                \end{array}\right]e^{in\theta}} \, ,
\end{equation}
where $f_{n,ss}=S_{n,ss}-1$, $f_{n,s\bar{s}}=i \bar{s}
S_{n,s\bar{s}}$, and the sum over $n$ in (\ref{fmatrix}) runs over
all integers. Conservation of flux for each channel of angular
momentum (unitarity of $S$), imposes the condition
$|S_{n,ss}|^2+|S_{n,s\bar{s}}|^2=1$, so that one can relate the
scattering amplitudes to the phase shifts gained during the
scattering process by $S_{n,ss}\equiv
e^{2i\delta_{n,ss}}\cos\delta_{n,s\bar{s}}$ and
$S_{n,s\bar{s}}\equiv \sin\delta_{n,s\bar{s}}$, where
$\delta_{n,ss}$ is the phase for spin preserving processes and
$\delta_{n,s\bar{s}}$ is conveniently defined for spin-flipping
events \cite{biref,mediateso,ellyot}. The description above, an
extension of the partial wave component method \cite{elastic},
allows for the exploration of spin-dependent phenomena \cite{biref}
and observables such as: the differential cross section
$\sigma(\theta)$, that explicitly displays the anisotropy of the
scattering; the transport cross section $\sigma_{tr}$, related to
the transport mean free time,
$\tau_{tr}^{-1}=n_{imp}v_{F}\sigma_{tr}$; and the total cross
section $\sigma_{t}$, related to the elastic scattering time,
$\tau_{e}^{-1}=n_{imp}v_{F}\sigma_{t}$, where $n_{imp}$ is the
impurity concentration in the sample. In the presence of SOIs the
scattering includes spin-preserving and spin-flip events.
Correspondingly, all these cross sections are spin-dependent
matrices given by

\begin{subequations}
\begin{equation}\label{diff}
\sigma_{ss'} \left(\theta\right)=\frac{1}{2\pi k}\left|\sum_{n}
{f_{n,ss'}\, e^{in\theta}}\right|^{2}\, ,
\end{equation}
\begin{equation}\label{total}
\sigma_{t,ss'}=\frac{1}{k}\sum_{n} {\left|f_{n,ss'}\right|^2} \, ,
\end{equation}
and
\begin{equation}\label{trans}
\sigma_{tr,ss'} =\sigma_{t,ss'}-\frac{1}{ k}\sum_{n} {{\rm
Re}\left(f_{n,ss'}f_{n+1,ss'}^{*}\right)}\, .
\end{equation}
\end{subequations}

In the absence of SOIs the pseudo-helicity,
$\mathbf{\sigma}\cdot\mathbf{p}/p$ is a conserved quantity
\cite{elctronincproperties,elastic} and results in the equality
$f_{m}\equiv f_{-(m-1)}$, where $m$ is an integer
($n=m\mp\frac{1}{2}$ for $\uparrow/\downarrow$) \cite{elastic,SuppInfo}, which leads to a
vanishing differential cross section at $\theta=\pi$ (Klein
tunneling), $\sigma(\theta=\pi)=0$, indicating the anisotropic
character of the scattering process and the near transparency of
barriers in graphene \cite{40s,Klein,Cheianov,gate3}. At low carrier
concentrations, $kR\ll1$, $f_{0}\equiv f_{1}$ and
$f_{m\neq0,1}\approx (kR)^m$, leading to $\sigma_{t}\simeq
2\sigma_{tr}$, and therefore $\xi=\tau_{tr}/\tau_{e}\simeq 2$.
Therefore, scattering of massless Dirac fermions in graphene from
short range potential scatterers results in $\xi\simeq 2$, for all
$V$ and $R$, as long as the carrier density is small, $kR\ll1$
\cite{helen}. This ratio is fully determined by the number and equal
weights of the angular momentum channels contributing to the
scattering process.  As we will see below, this situation is
drastically changed in the presence of SOI.

\emph{\textbf{Graphene with intrinsic SOI.}} Graphene systems with
uniform intrinsic SOI (for space dependent $\Delta_{SO}$ see
\cite{SuppInfo}), $\Delta_{SO} \neq 0$, represent a rich opportunity
to explore topological effects. An example of such a system is
predicted by appropriate deposition of heavy metal atoms on graphene
\cite{indium}. In those cases, the eigenstates no longer have a
well-defined pseudo-helicity, due to the carrier mass generated by
the SOI; notice however that although this mass is spin-dependent,
it does {\em not} cause intravalley spin-flip processes, and the
scattering can still be analyzed in terms of independent spins. The
broken pseudo-helicity, however, results in
$\delta_{n,ss}\neq\delta_{-(n-1),ss}$. However, effective time
reversal symmetry \cite{BeenakkerRMP} imposes the relations
$f_{n,ss}=f_{-n,\bar{s}\bar{s}}$, and
$f_{n,s\bar{s}}=f_{-n,\bar{s}s}$, and since spin mixing is not
produced by the intrinsic SOI, we have
$\delta_{n,s\bar{s}}=\delta_{-n,\bar{s}s}=0$.

\begin{figure}
\includegraphics[scale=0.32]{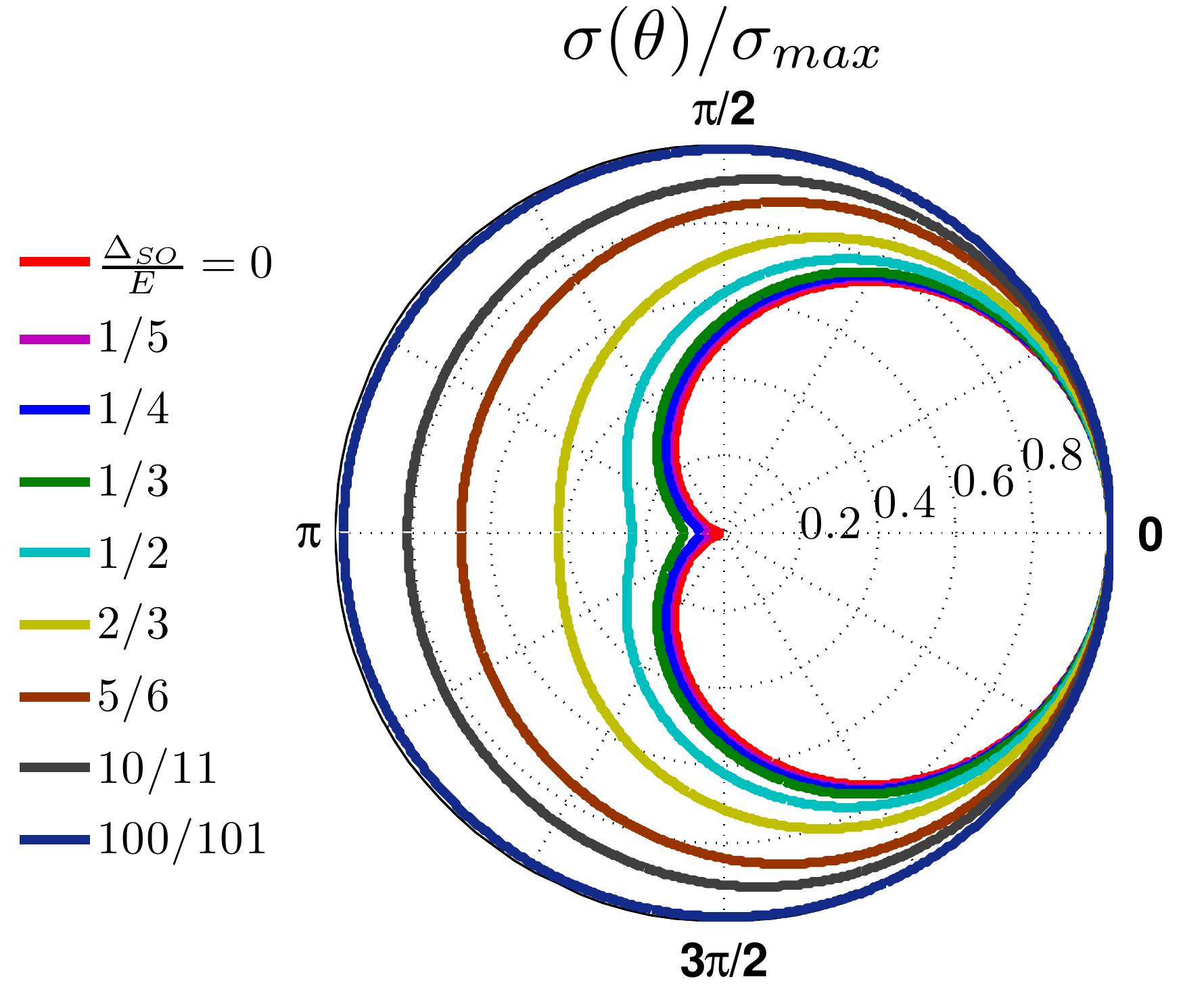}
\includegraphics[scale=0.31]{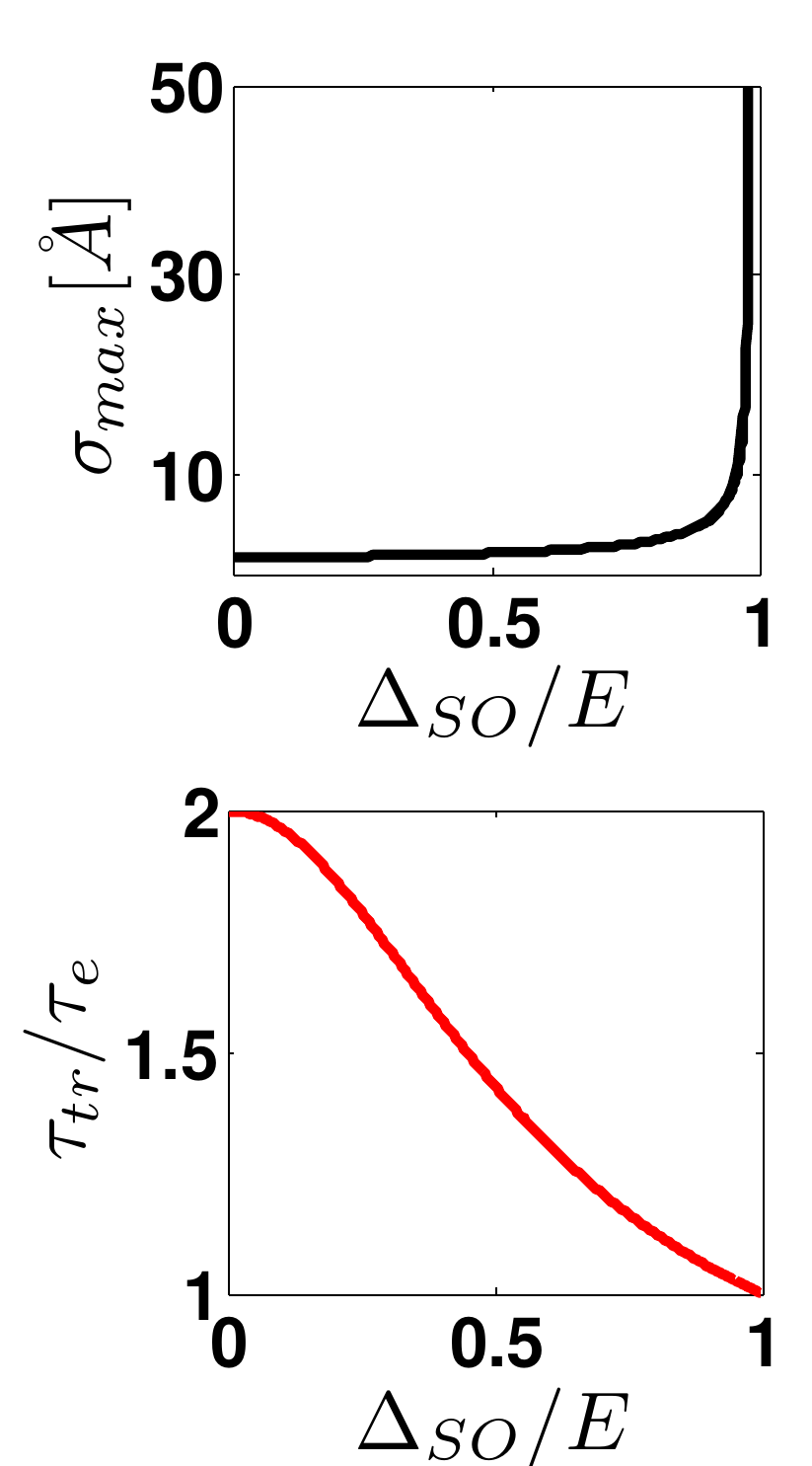}
  \caption{
 Polar plots of differential cross section, normalized to its
maximum, for different values of the intrinsic spin orbit
interaction, $\Delta_{SO}$; here $ER/(\hbar v_{F})=8\times10^{-3}$
and $VR/(\hbar v_{F})=1.5$. Top inset: $\sigma_{max}(\theta)$, which
increases as $1/(k\ln^2(kR))$ for $\Delta_{SO}/E\approx1$, sets the
scale used in the polar plots. Bottom inset: Dependence of
$\xi=\sigma_{t}/\sigma_{tr}=\tau_{tr}/\tau_{e}$ vs.\
$\Delta_{SO}/E$. Notice that $\sigma(\theta)/\sigma_{max}$
and $\xi$ do not depend on the value of $V$ in this regime; $V$ only determines
the amplitude of $\sigma_{max}$ in the top inset.}
  \label{Fig1}
\end{figure}

As one could suspect, the isotropy of the scattering process depends
on the ratio of $\Delta_{SO}/E$, as shown in Fig.\ \ref{Fig1}: the
scattering is anisotropic--with absence of back scattering--for
$\Delta_{SO}=0$, while it becomes increasingly isotropic with larger
 $\Delta_{SO} /E$, and for $\Delta_{SO}\approx E$, the scattering
is equally probable in all directions.

The change in the isotropy of the scattering process is related to
the total number of angular momentum channels contributing to the
cross section. For an incoming electron flux with ``high'' energy,
$0\leq \Delta_{SO}/E\ll 1$, the system exhibits approximately equal
contributions from two scattering channels, $n=0$ and $n=1$ for
$\uparrow$ incoming flux (or $n=0$ and $n=-1$ for $\downarrow$
incident flux), and these contributions satisfy $f_{0,\uparrow
\uparrow}\approx f_{1,\uparrow \uparrow}$ (or $f_{0,\downarrow
\downarrow} \approx f_{-1,\downarrow \downarrow}$). In contrast, we
observe an increase in the isotropy of the scattering as $E$
decreases, approaching $\Delta_{SO}$, due to the vanishing
contribution of the $n=0$ channel to the total cross section,
$\vartheta(k^3R^4)$, compared to resonant contribution of the
$n=\pm1$ channels $\pi^2/(k\ln^2(kR))$ \cite{SuppInfo}. This leads
to the ``isotropic'' ratio of $\xi=\tau_{tr}/\tau_{e}\approx 1$,
which is characteristic of the scattering of massive particles at
low energies; in other words, one of the spinor components dominates
the scattering process in this range of energy and leads to a fully
isotropic differential scattering cross section. As $\Delta_{SO}$
determines the energy scale for which the isotropy would play a
larger role, the exploration of decorated graphene samples would be
an interesting system in which to test these results \cite{indium}.

\emph{\textbf{Graphene with Rashba SOI.}}  We now analyze the case
of graphene samples containing scattering centers that also produce
Rashba interactions \cite{impurityso,Nii,golddep,colossal},
$\Delta_{SO}\ll\lambda_R \neq 0$~\cite{SuppInfo}, allowing spin flip
events. This requires a detailed analysis of the spin dependent
scattering processes. When $kR\ll 1$, we have two contributing
channels, depending on the spin of the incoming particle ($n=0,1$
for spin up, and $n=0,-1$ for spin down), similar to the case
discussed above for $\Delta_{SO}\neq 0$. Effective time reversal
symmetry within the Dirac cone allows one to study the scattering of
a given spin without loss of generality
\cite{BeenakkerRMP,SuppInfo}.

\begin{figure}
\centering
 \includegraphics[scale=0.2]{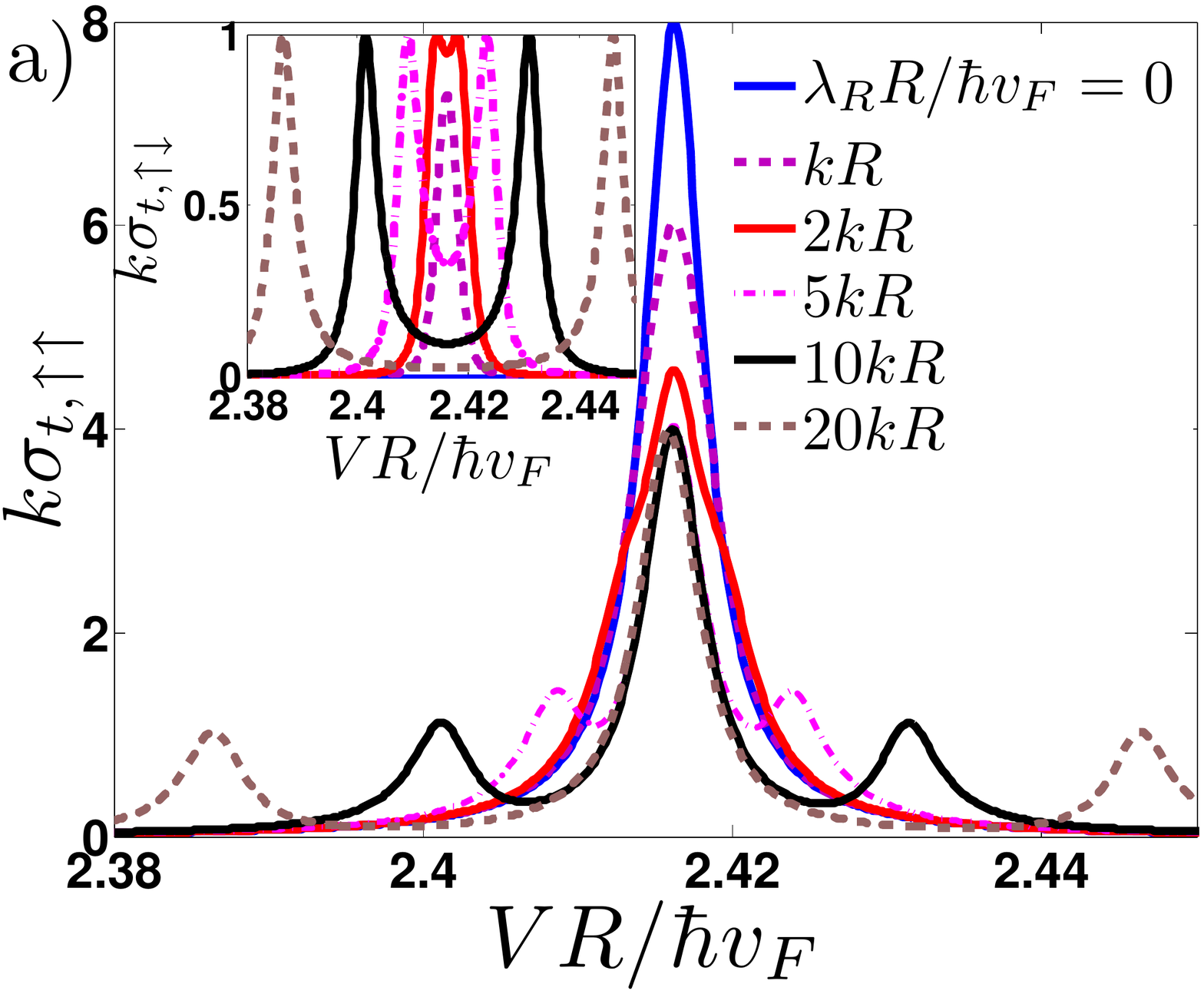}
 \includegraphics[scale=0.2]{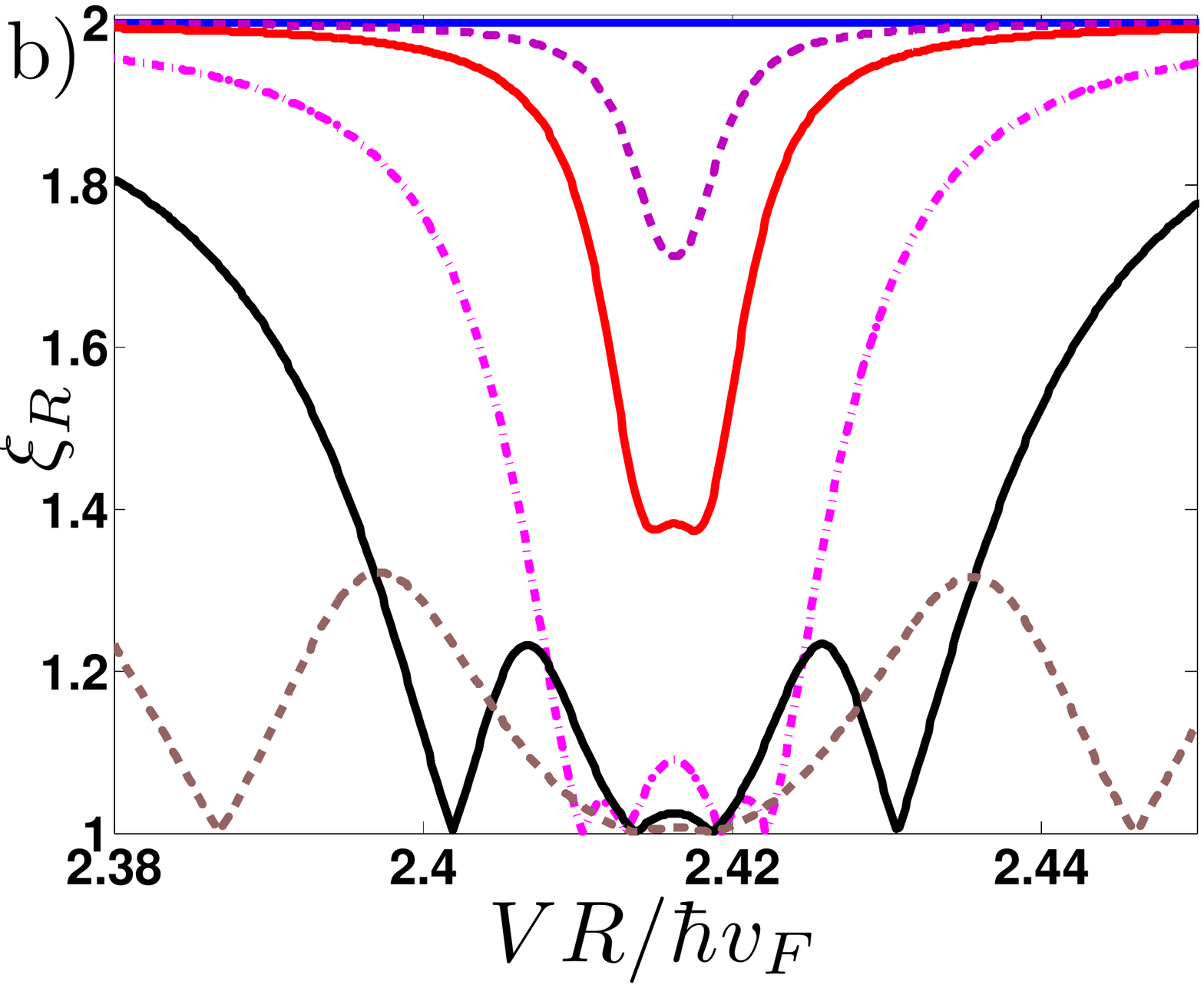}
  \caption{ a) Total cross section for spin-preserving processes as function of the scattering potential shift $V$,
  for different values of the Rashba SOI,
   with $kR=1.5\times10^{-3}$ and $\Delta_{SO}=0$ (for $\Delta_{SO}\ne0$ see \cite{SuppInfo}). Inset: Total cross section for
spin-flip processes. b) The ratio
   $\xi_{R}=(\sigma_{t,\uparrow\uparrow}+\sigma_{t,\uparrow\downarrow})/(\sigma_{tr,\uparrow\uparrow}+\sigma_{tr,\uparrow\downarrow})$
   for different values of Rashba SOI (legend as in a). Notice $\xi_R =1$ at $\sigma_t$ resonances.}
  \label{Fig2}
\end{figure}

Curves of total cross section vs.\ scattering potential strength $V$
are shown in Fig.\ \ref{Fig2}a for $kR\ll1$, and different values of
the Rashba SOI interaction, $\lambda_R$; analytical expressions for
the different contributions can be obtained as well \cite{SuppInfo}.
Figure \ref{Fig2}a shows how the location and number of resonances
change in the presence of Rashba SOIs.  The resonances at
$\chi'=\chi_{0}\pm\lambda_R R/\hbar v_{F}$ for both $\sigma_{t,ss}$
and $\sigma_{t,s\bar{s}}$, can be identified as resonances of the
$n=0$ channel, while the resonance at $\chi'\approx
\chi_{0}+\vartheta((\lambda_{R}R/(\hbar v_{F}))^2)$ can be
identified as coming from the $n=1$ ($-1$) for $s=\uparrow$
($\downarrow$) incoming spin, where $\chi_{0}$ is the location  of
the unitary resonance in the absence of SOIs. Similarly, Fig.\
\ref{Fig2}b demonstrates that the scattering isotropy at resonant
values is different from the case of no SOI, by showing that the
ratio $\xi_R=(\sigma_{t,\uparrow\uparrow}+\sigma_{t,\uparrow\downarrow})%
/(\sigma_{tr,\uparrow\uparrow}+\sigma_{tr,\uparrow\downarrow})$
takes on different values in the different regimes, being
$\xi_R\simeq1$ for unitary resonances, $1<\xi_R<2$ for medium
scatterers, and $\xi_R \simeq2$ when off-resonance. This qualitative
difference arises from the fact that the scattering amplitudes of
the two contributing channels are not equal for all the scattering
regimes, in contrast to the case of scattering in the absence of the
Rashba interaction, where $\xi\simeq2$ for all regimes, off- and
on-resonance.

\begin{figure}
\centering
 \includegraphics[scale=0.3]{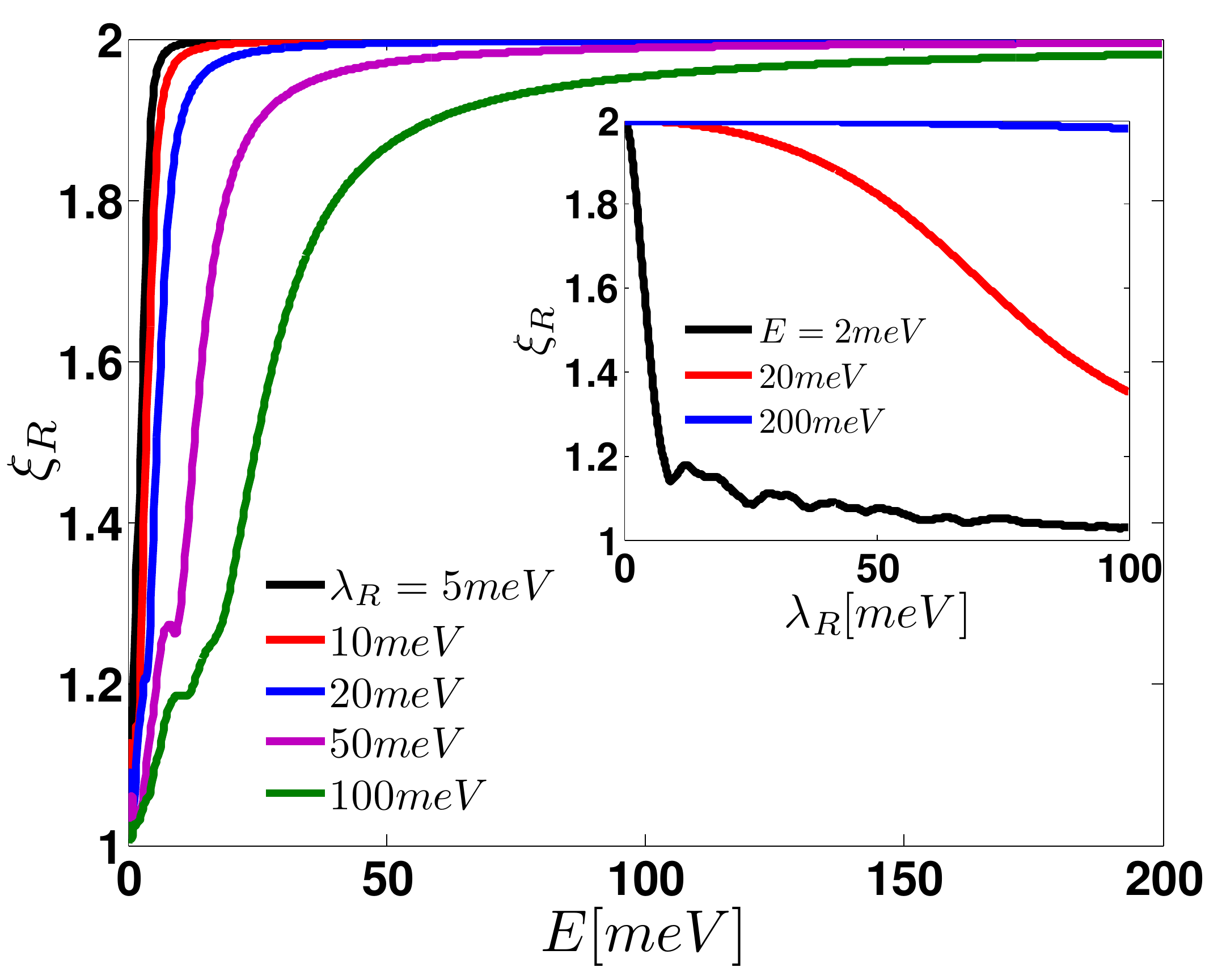}
  \caption{  The ratio $\xi_{R}$ for $500$ randomly sized impurities in the range of
5\AA $\,\leq R\leq 8$\AA, for different values of the Rashba
coupling and $V=2 eV$, as a function
   of carrier energy. Notice a clear drop of $\xi_R$ from $2$ for $E<\lambda_{R}/2$.
   Inset: $\xi_{R}$ as function of Rashba coupling for different energies.}
  \label{Fig3}
\end{figure}

To further explore the consequences of this SOI-dependent behavior
on transport experiments \cite{helen}, we consider a random
distribution of scatterers in a typical graphene sample.  The
distribution is assumed to be of low-density, as we ignore multiple
scattering events. Moreover, as the parameter in the theory is $VR$,
we assume a random distribution for that quantity in the range $1.5$
to $2.4$ (in units of $\hbar v_F$). For a fixed value $V\simeq2$eV,
for example, this would correspond to a variation in $R$ from
$\simeq5$ to 8\AA, not unlike those considered before
\cite{impurityso,helen}.
The results of such averaging procedure are shown in Fig.\
\ref{Fig3}, where $\xi_{R} = \left<\sigma_{t,\uparrow \uparrow} +
\sigma_{t,\uparrow \downarrow}\right> /\left<\sigma_{tr,\uparrow
\uparrow} + \sigma_{tr,\uparrow \downarrow}\right>$ is shown as
function of (Fermi) energy for different values of the Rashba SOI
strength $\lambda_R$, while $V=2$eV is kept fixed. Notice that the
range of $E$ in the figure satisfies $kR < 0.24$ for all values
shown and can therefore be understood in terms of the analytical
expansions above--however, the curves shown are obtained from a full
numerical evaluation of the different cross sections that consider
multiple channels. As one would expect, as the energy (or carrier
density) increases, the ratio $\xi_{R}$ approaches the anisotropic,
effectively SOI-free limit, $\approx 2$, while at low energies,
$\xi_R$ approaches the isotropic scattering limit of 1. The drop
occurs for a characteristic energy given by $\lambda_R$, with
$\xi_{R} \simeq 1.8$ for
 $E \simeq \lambda_R/2$; this condition can be traced back  to the
 shifting resonances of the $n=0$ channel under Rashba SOI.
One can also analyze the dependence of $\xi_{R}$ on the Rashba
coupling for different carrier densities (energies), as shown in the
inset of Fig.\ \ref{Fig3}.  It is evident that the effect of a small
Rashba coupling is more pronounced at lower energies.

From the preceding analysis, it appears that the experimental
evaluation of the transport to elastic times ratio at low carrier
densities would be able to provide an alternative measure of the
effective Rashba SOI present, as produced by impurities and defects,
either intrinsic or purposely introduced. Such careful experiments
have already explored this ratio \cite{helen}, and as the carrier
density has been reduced down to $E\approx 100$meV, it appears the
induced Rashba SOI in those samples was well below that number
(i.e., $\lambda_R < 200$meV), since $\xi_R \simeq 2$ over the entire
range explored.  We believe it would be interesting to repeat those
experiments in systems with higher mobility, such as graphene on
boron nitride substrates, which may allow reaching even lower
carrier densities without large inhomogeneities.  Considering that
in systems with adatoms the expected SOI is $\lambda_R \approx
10$meV \cite{impurityso,colossal,golddep}, this requires rather low
carrier densities, such as those attained on boron nitride
substrates \cite{diracpoint1,ranom}.

We should comment that the observed renormalization of the Fermi
velocity near the Diract point \cite{diracpoint1,ranom} which sees
the velocity increase as the energy (or carrier density) drops,
should result in $\xi_R$ dropping down from 2 at a higher energy
than in the absence of the velocity renormalization (for a given
$\lambda_R$, and assuming a large enough $VR$, so that $v_F$
rescaling at $V$ is negligible).

In conclusion, we have shown that SOIs in graphene lead to clear
signatures in the scattering processes and therefore to observable
consequences in electronic transport. The drop in value of the ratio
of transport to elastic times from its known value of $\simeq 2$
reflects the presence of SOI, with the ratio dropping to $\simeq 1$
as $E_F$ falls close to the SOI energy scale. We have also shown
qualitative changes in the number and nature of resonances produced
in scattering due to impurities and the Rashba SOI they induce.
Three different regimes of scattering can be distinguished based on
the levels of isotropy they produce, with the isotropy becoming more
pronounced at low carrier concentrations. Measuring the ratio of
scattering times with precision at low carrier densities should
enable the experimental characterization of impurity-induced
spin-orbit interactions.

We thank N. Sandler and M. Zarea for helpful discussions. This work
was supported in part by NSF PIRE, and NSF CIAM/MWN grant
DMR-1108285. We are grateful for the welcoming environment at the
Dahlem Center and the support of the A. von Humboldt Foundation.

\bibliography{referencessoi}

\end{document}